\documentclass[10pt,twocolumn,letterpaper]{article}
\usepackage{cvpr} 
\usepackage{algorithmic}
\usepackage{algorithm}
\usepackage{multirow}
\usepackage[table]{xcolor}
\definecolor{verylightgray}{gray}{0.9}

\newcommand\shline{\specialrule{0.8pt}{0pt}{0pt}}
\definecolor{cvprblue}{rgb}{0.21,0.49,0.74}
\usepackage[pagebackref,breaklinks,colorlinks,allcolors=cvprblue]{hyperref}

\def\paperID{*****}
\def\confName{CVPR}
\def\confYear{2025}

\title{OSMamba: Omnidirectional Spectral Mamba with Dual-Domain Prior Generator for Exposure Correction}

\author{
{Gehui Li}$^{1,*}$ \qquad
{Bin Chen}$^{1,*}$ \qquad
{Chen Zhao}$^{3,\dag}$ \qquad
{Lei Zhang}$^{4, 5}$ \qquad
{Jian Zhang}$^{1, 2}$\\
$^1$School of Electronic and Computer Engineering, Peking University \\
$^2$Guangdong Provincial Key Laboratory of Ultra High Definition Immersive Media Technology,\\
Shenzhen Graduate School, Peking University\\
$^3$King Abdullah University of Science and Technology \\
$^4$The Hong Kong Polytechnic University \qquad 
$^5$OPPO Research Institute\\
{\tt\footnotesize \{ligehui921, chenbin\}@stu.pku.edu.cn} \qquad
{\tt\footnotesize chen.zhao@kaust.edu.sa} \\
{\tt\footnotesize cslzhang@comp.polyu.edu.hk} \qquad
{\tt\footnotesize zhangjian.sz@pku.edu.cn}}

\begin{document}
\maketitle

\begin{abstract}
Exposure correction is a fundamental problem in computer vision and image processing. Recently, frequency domain-based methods have achieved impressive improvement, yet they still struggle with complex real-world scenarios under extreme exposure conditions. This is due to the local convolutional receptive fields failing to model long-range dependencies in the spectrum, and the non-generative learning paradigm being inadequate for retrieving lost details from severely degraded regions. In this paper, we propose \textbf{O}mnidirectional \textbf{S}pectral \textbf{Mamba} (\textbf{OSMamba}), a novel exposure correction network that incorporates the advantages of state space models and generative diffusion models to address these limitations. Specifically, OSMamba introduces an omnidirectional spectral scanning mechanism that adapts Mamba to the frequency domain to capture comprehensive long-range dependencies in both the amplitude and phase spectra of deep image features, hence enhancing illumination correction and structure recovery. Furthermore, we develop a dual-domain prior generator that learns from well-exposed images to generate a degradation-free diffusion prior containing correct information about severely under- and over-exposed regions for better detail restoration. Extensive experiments on multiple-exposure and mixed-exposure datasets demonstrate that the proposed OSMamba achieves state-of-the-art performance both quantitatively and qualitatively.
\end{abstract}

\renewcommand{\thefootnote}{}
\footnote{$^*$Equal Contribution. $^\dagger$Corresponding Author.}

\vspace{-30pt}
\section{Introduction}
In real-world scenarios, images often suffer from under- or over-exposure due to complex and variable lighting conditions, inherent limitations of capturing devices, and potential operator errors~\cite{afifi2021msec,bychkovsky2011learning,liu2024deep}. For example, under-exposure occurs when image details are obscured in shadows, significantly reducing contrast and clarity; over-exposure can lead to saturation in highlight areas, resulting in irreversible loss of important details. These issues severely affect image quality and can also impact many applications including but not limited to visual recognition~\cite{cho2024dual, onzon2024neural, kim2024beyond} and video processing~\cite{jin2024chat, chang2024towards, xie2024uveb}. Consequently, exposure correction~\cite{afifi2021msec, wang2022lcdp, cai2018learning} has been a critical problem in computer vision, attracting widespread attention from researchers. 

\begin{figure}[!t]
\centering
\includegraphics[width=0.93\columnwidth]{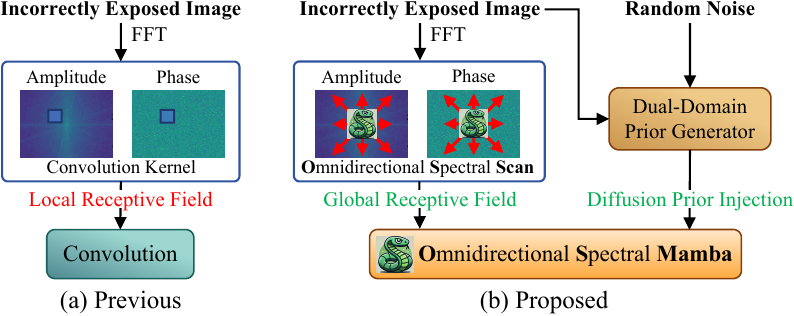}
\vspace{-7pt}
\caption{\textbf{Comparison of OSMamba with previous methods.} \textbf{(a)} Previous frequency-domain approaches (e.g., \cite{huang2022fecnet}) process spectrum with convolutions, limited by local receptive fields. \textbf{(b)} Proposed OSMamba enjoys global receptive field by using omnidirectional spectral scanning. It also leverages a dual-domain prior generator to generate and inject diffusion prior into the network.}
\label{fig:contribution}
\vspace{-13pt}
\end{figure}

In the past decade, many low-light enhancement methods~\cite{chen2018learning, wang2019underexposed, wei2018deep, ma2022toward, yang2023implicit, yang2023difflle} have been developed. They usually assume that input images are purely under-exposed and employ the Retinex theory~\cite{rahman2004retinex} to decompose images into illumination and reflectance components for separate processing. However, these methods falter when faced with multi-exposure conditions. The emergence of deep networks \cite{ronneberger2015u,zamir2022restormer,liu2024compressive,chen2025invertible,chen2024self,chen2024practical} has changed the field of exposure correction by enabling end-to-end learning for directly predicting corrected results from inputs. Notably, recent methods~\cite{afifi2021msec, wang2022lcdp, huang2022fecnet, baek2023lact, li2024real} incorporate classical image processing algorithms into deep networks for improved performance.

Nevertheless, these methods are hindered by two drawbacks in real-world scenarios with extreme exposure conditions. Firstly, they face challenges in illumination correction and structure recovery. Recent advanced frequency domain-based approaches like FECNet~\cite{huang2022fecnet} have demonstrated greater effectiveness than previous spatial domain-based methods. This superiority stems from leveraging the well-decoupled properties of Fourier transform to separately optimize amplitude and phase spectra, which represent illumination and structure components respectively. However, their performance is constrained by relying solely on small-sized convolution kernels. As Figure~\ref{fig:contribution}(a) shows, these convolution kernels have local receptive fields, leading to limited representation ability for modeling long-range dependencies among elements in the amplitude and phase spectra. Secondly, they struggle to restore lost details in extremely under- and over-exposed regions. This can be attributed to the fact that they only train discriminative networks using regression losses (e.g. $L_1$ and $L_2$ losses) to learn the mapping from exposure-error images to ground-truth images. When inputs undergo severe degradation, high-frequency details are difficult to retrieve through regression-based discriminative learning~\cite{saharia2022image, wu2023latent}.

\begin{figure}[!t]
\centering
\includegraphics[width=0.85\columnwidth]{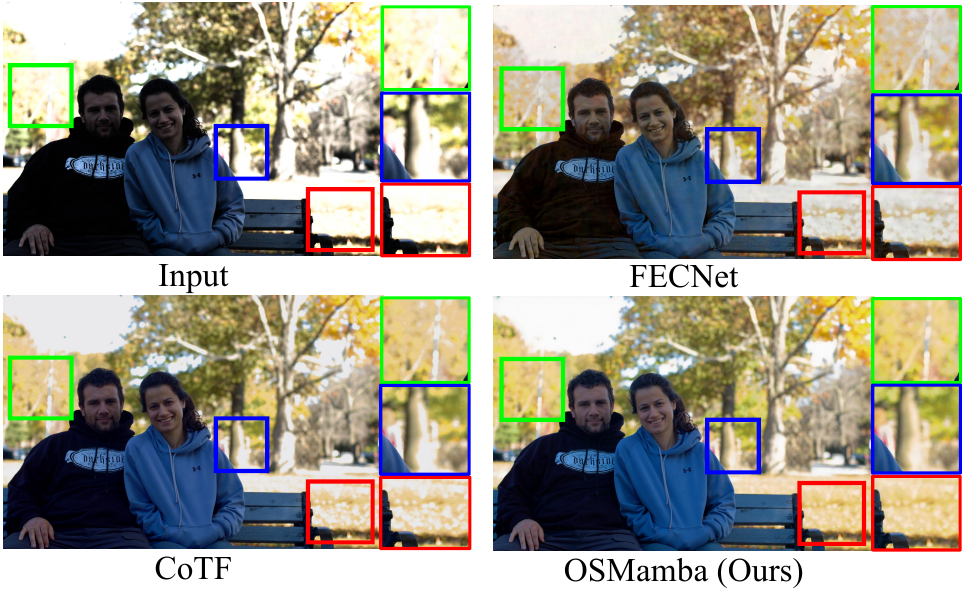}
\vspace{-10pt}
\caption{\textbf{Visual comparison of proposed OSMamba with state-of-the-arts.} The green, blue, and red boxes highlight our method's advantages over FECNet~\cite{huang2022fecnet} and COTF~\cite{li2024real} in correction quality.}
\label{fig:contrast}
\vspace{-13pt}
\end{figure}

To more effectively capture the long-range feature dependencies in the frequency spectrum and overcome the inability of non-generative networks to restore lost details, we propose \textbf{O}mnidirectional \textbf{S}pectral \textbf{Mamba} (\textbf{OSMamba}) for exposure correction. Our approach is motivated by the recent success of State Space Models (SSMs)~\cite{gu2021efficiently, gu2023mamba, dao2024transformers} in image processing, where they have demonstrated great efficacy in global modeling with linear complexity to sequence length. However, when directly applied to the frequency domain, the scanning mechanisms of existing SSM-based networks~\cite{zhu2024vision, liu2024vmamba, guo2024mambair} face two major limitations: \textbf{(1)} their row- and column-wise scanning mechanisms fail to capture diagonal frequency interactions in the spectrum, and \textbf{(2)} they do not fully exploit the inherent properties of frequency spectrum such as central symmetry and continuity. To this end, OSMamba introduces a novel Omnidirectional Spectral Scanning (OS-Scan) mechanism that incorporates four continuous scanning methods: row-wise, column-wise, positive diagonal, and negative diagonal scanning for processing half of the amplitude and phase features. This design improves the diversity of spectral interactions by capturing dependencies along rows, columns, and diagonals, as shown in Figure~\ref{fig:contribution}~(b), thereby improving the network's ability to correct illumination and recover structure.

Furthermore, to better restore the lost details, we leverage the powerful generation capabilities of diffusion models~\cite{xia2023diffir} and design a Dual-Domain Prior Generator (DDPG). As illustrated in Figure~\ref{fig:contribution}~(b), DDPG aggregates spatial and frequency domain information from the input image as the condition, while using sampled noise as the starting point, to generate a degradation-free prior that contains correct information about severely degraded regions. Figure~\ref{fig:contrast} demonstrates our method's superiority over previous state-of-the-art approaches in correction quality.

To summarize, our contributions are three-fold:
\begin{itemize}
\item We propose a novel exposure correction method OSMamba, which introduces Mamba with an Omnidirectional Spectral Scanning mechanism to effectively model comprehensive long-range dependencies within amplitude and phase spectra, leading to superior illumination correction and structure recovery.

\item We develop a Dual-Domain Prior Generator that learns from well-exposed images to generate a degradation-free diffusion prior containing correct information about severely degraded regions for better detail restoration.

\item Experiments on three exposure correction benchmarks demonstrate that our OSMamba achieves state-of-the-art performance both quantitatively and qualitatively.
\end{itemize}

\section{Related Work}
\subsection{Exposure Correction}  
Early works focused on addressing under- or over-exposure using a unified framework, exemplified by MSECNet~\cite{afifi2021msec}, which adopted a coarse-to-fine network architecture. Later research further tried to correct images with mixed exposure. For example, LCDPNet~\cite{wang2022lcdp} introduced the concept of local color distributions to correct under- and over-exposed regions within a single input image.

A critical challenge in exposure correction is maintaining consistent outputs across inputs with different exposures. Several works have proposed innovative solutions to this problem~\cite{huang2022enc,huang2022eclnet,huang2023erl,li2023fearless}. ERL~\cite{huang2023erl} explored the relationship between under- and over-exposure optimization by correlating and constraining the correction procedures within a mini-batch. Recent works~\cite{cui2022iat,huang2022fecnet,baek2023lact,liu2024region,li2024real} have also explored new methodologies for exposure correction. LACT~\cite{baek2023lact} introduced an illumination-aware color transform algorithm, focusing on preserving image details. CoTF~\cite{li2024real} combined image-adaptive 3D LUTs for global adjustments with pixel-wise transformations for local refinement. 

\subsection{State Space Model}  
Several works~\cite{zhu2024vision,liu2024vmamba,shaker2024groupmamba, zhao2024rs} have been devoted to developing new vision backbones based on State Space Models (SSMs)~\cite{gu2021efficiently, gu2023mamba}. For example, VMamba~\cite{liu2024vmamba} proposed Visual State Space (VSS) blocks with Cross-Scan for efficient global modeling. RS-Mamba~\cite{zhao2024rs} employed an omnidirectional selective scan module, effectively capturing context across various spatial directions in the tasks of remote sensing dense prediction. ZigMa~\cite{hu2024zigma} combined Mamba with a Zig-Zag scanning scheme in diffusion models, improving the scalability for high-resolution datasets. Researchers have also explored integrating SSMs with image restoration tasks~\cite{shi2024vmambair,guo2024mambair,li2024fouriermamba,zhen2024freqmamba, zou2024wave, wu2024rainmamba, dong2024ecmamba}. FourierMamba~\cite{li2024fouriermamba} introduced frequency scanning in Fourier domain for image deraining, while MambaIR~\cite{guo2024mambair} enhanced vanilla Mamba with local modeling and channel attention for image super-resolution and denoising. 
Moving beyond these works, we propose OSMamba to establish diverse long-range dependencies in the frequency domain through OS-Scan for better illumination correction and structure recovery. Additionally, we develop a DDPG to guide OSMamba to restore details under extreme exposure conditions.

\begin{figure*}[!t]
\vspace{-10pt}
\centering
\includegraphics[width=0.99\textwidth]{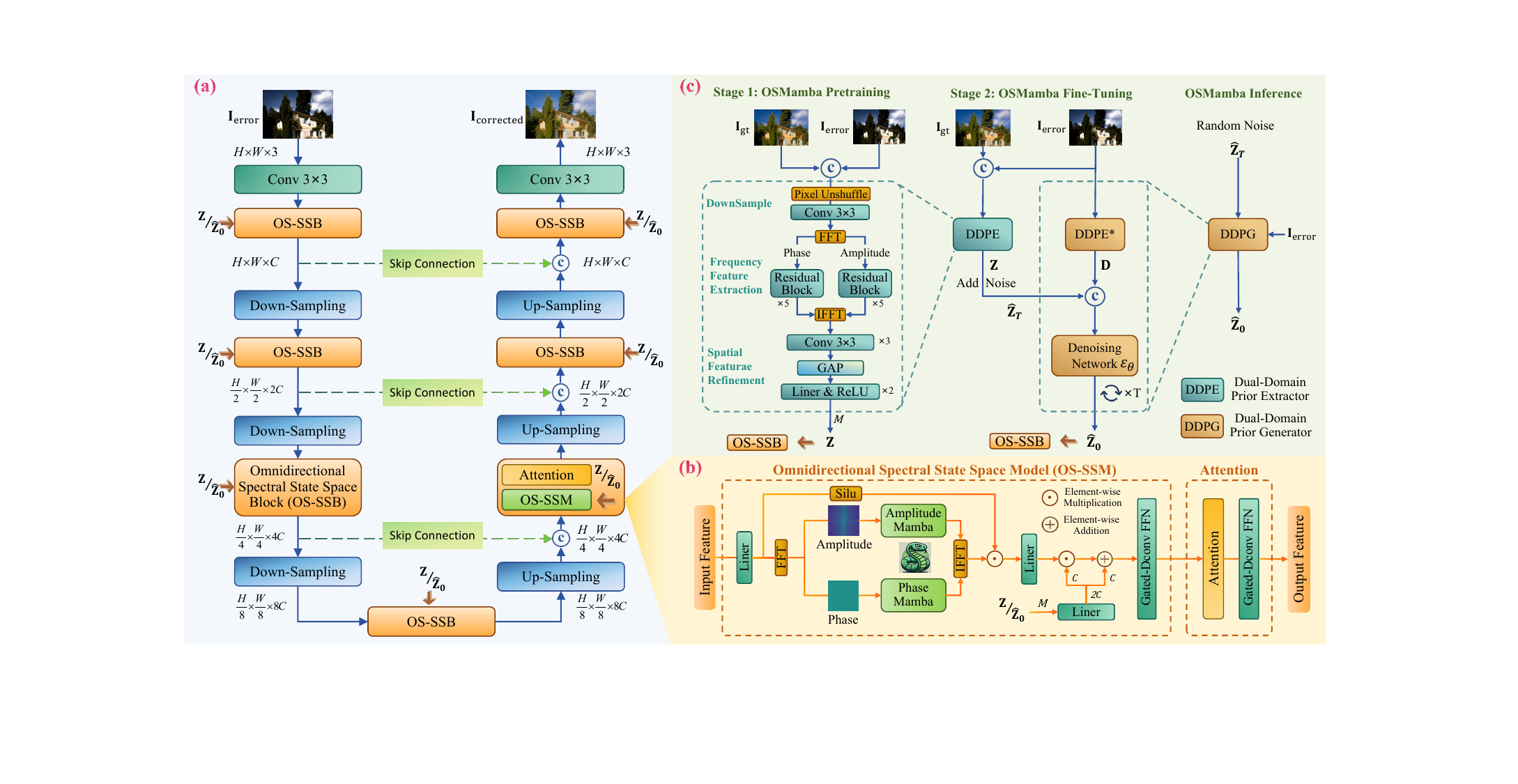}
\vspace{-5pt}
\caption{\textbf{Illustration of proposed OSMamba.} \textbf{(a)} The overall architecture of OSMamba consists of multiple levels of Omnidirectional Spectral State Space Block (OS-SSB). \textbf{(b)} Each OS-SSB contains an Omnidirectional Spectral SSM (OS-SSM) and an Attention Module. The OS-SSM performs Omnidirectional Spectral Scanning (OS-Scan) on the amplitude and phase of deep features using Amplitude Mamba and Phase Mamba. \textbf{(c)} The compact teacher prior $\mathbf{Z}$ extracted by Dual-Domain Prior Extractor (DDPE) and the diffusion prior $\mathbf{\hat{Z}}_{0}$ generated by Dual-Domain Prior Generator (DDPG) are input to guide all OS-SSMs in Stages 1 (Pretraining) and 2 (Finetuning), respectively. During inference, $\mathbf{Z}$ is replaced by a randomly sampled Gaussian noise, and $\mathbf{\hat{Z}}_{0}$ is generated based on a condition $\mathbf{D}$ to guide detail restoration.}
\label{fig:structure}
\vspace{-12pt}
\end{figure*}

\section{Proposed Method}
\subsection{Preliminary}
\noindent \textbf{State Space Model.} The structured state space sequence models (S4)~\cite{gu2021efficiently} have driven significant advancements in SSMs~\cite{han2024demystify, bick2024transformers}. Mamba (S6)~\cite{gu2023mamba} further utilizes a discretized version of a linear time-varying system that maps an input sequence $\mathbf{x} \in \mathbb{R}^{L \times C}$ to an output sequence $\mathbf{y} \in \mathbb{R}^{L \times C}$ with the same length through a latent state $\mathbf{h}_k \in \mathbb{R}^{C \times N}$. The formulation of S6 can be expressed as:
\begin{equation}
\begin{aligned}
\label{eq:discret-ssm}
\mathbf{h}_k &= \overline{\rm \mathbf{A}} \mathbf{h}_{k-1} + \overline{\rm \mathbf{B}}\mathbf{x}_k, \\
\mathbf{y}_k &= {\rm \mathbf{C}} \mathbf{h}_k,
\end{aligned}
\end{equation}
where $\overline{\rm \mathbf{A}} \in \mathbb{R}^{L \times C \times N}$, $\overline{\rm \mathbf{B}} \in \mathbb{R}^{L \times C \times N}$ and ${\rm \mathbf{C}} \in \mathbb{R}^{L \times N}$ with $L$, $C$, $N$ denoting sequence length, number of channels and state size. The discrete $\overline{\rm \mathbf{A}}$ and $\overline{\rm \mathbf{B}}$ are obtained from ${\rm \mathbf{A}}$ and ${\rm \mathbf{B}}$ through zero-order hold discretization~\cite{gu2021efficiently}:
\begin{equation}
\begin{aligned}
\overline{\rm \mathbf{A}} &= {\rm exp}({\rm {\Delta \mathbf{A}}}), \\
\overline{\rm \mathbf{B}} &= ({\rm {\Delta \mathbf{A}}})^{-1}({\rm exp(\mathbf{A})}-\mathbf{I}) \cdot {\rm \Delta \mathbf{B}} \approx {\rm \Delta \mathbf{B}},
\end{aligned}
\end{equation}
where $\Delta$ is the timescale parameter to control the discretization level. This formulation allows Mamba to efficiently process 1D data while capturing long-range dependencies with computational complexity linear to the sequence length. The scanning mechanism~\cite{huang2024localmamba, liu2024vmamba, hu2024zigma} unfolds 2D feature maps into 1D sequences, which makes Mamba suitable for efficient global modeling of deep image features in exposure correction.

\noindent \textbf{2D Discrete Fourier Transform.} For an image $\mathbf{X}(h,w)$ of size $H \times W$, its 2D Fourier transform $\mathcal{F}(\mathbf{X})$ is defined as:
\begin{equation}
\mathcal{F}(\mathbf{X})(u,v) = \sum_{h=0}^{H-1} \sum_{w=0}^{W-1} \mathbf{X}(h,w) e^{-j2\pi(\frac{uh}{H} + \frac{vw}{W})},
\end{equation}
where $\mathcal{F}(\mathbf{X})(u,v)$ is the result of the image's Fourier transform, and $(h,w)$ and $(u,v)$ represent spatial and frequency coordinates, respectively. This transformation can be efficiently computed on modern GPUs using the Fast Fourier Transform (FFT) algorithm~\cite{frigo1998fftw}. Based on  $\mathcal{F}(\mathbf{X})$, we can obtain the amplitude $\mathcal{A}(\mathbf{X})$ and the phase $\mathcal{P}(\mathbf{X})$: 
\begin{equation}
\label{eq:fft}
\begin{aligned}
\mathcal{A}(\mathbf{X})(u,v) &= \sqrt{\text{Re}(\mathcal{F}(\mathbf{X})(u,v))^2 + \text{Im}(\mathcal{F}(\mathbf{X})(u,v))^2}, \\
\mathcal{P}(\mathbf{X})(u,v) &= \text{arctan}\left(\frac{\text{Im}(\mathcal{F}(\mathbf{X})(u,v))}{\text{Re}(\mathcal{F}(\mathbf{X})(u,v))}\right),
\end{aligned}
\end{equation}
where $\text{Re}(\mathcal{F}(\mathbf{X})(u,v))$ and $\text{Im}(\mathcal{F}(\mathbf{X})(u,v))$ are the real and imaginary parts of $\mathcal{F}(\mathbf{X})(u,v)$, respectively. The amplitude spectrum $\mathcal{A}(\mathbf{X})$ represents the magnitude of frequency components, which correlates with the image's illumination intensity, while the phase spectrum $\mathcal{P}(\mathbf{X})$ encodes the positions of frequency components, crucial for preserving the image's structure information.

\subsection{Overall Framework of OSMamba}  
As illustrated in Figure~\ref{fig:structure}(a), given an exposure-error image $\mathbf{I}_\text{error} \in \mathbb{R}^{H\times W\times 3}$, we first use a convolution layer to extract a $C$-channel shallow feature. Then, we employ a four-level UNet~\cite{ronneberger2015u} to refine the feature with channel numbers  $C$, $2C$, $4C$, and $8C$, respectively. Each level incorporates an Omnidirectional Spectral State Space Block (OS-SSB), which consists of an Omnidirectional Spectral State Space Model (OS-SSM) and an Attention module~\cite{zamir2022restormer}, both followed by a Gated-DConv Feed-forward Network (GDFN)~\cite{zamir2022restormer} to enhance the inter-channel correlations. The corrected image $\mathbf{I}_\text{corrected}$ with the same size as the input $\mathbf{I}_\text{error}$ and ground truth $\mathbf{I}_\text{gt}$ is finally obtained through another convolution. To better restore lost details, we modulate the deep features in OS-SSM  by a diffusion prior $\mathbf{Z}_{0}\in\mathbb{R}^{1\times 1\times M}$ generated by Dual-Domain Prior Generator (DDPG). In the following three subsections, we will elaborate on the details of OS-SSM and how the prior is generated to guide OS-SSMs.

\begin{figure}[!t]
\centering
\includegraphics[width=0.9\columnwidth]{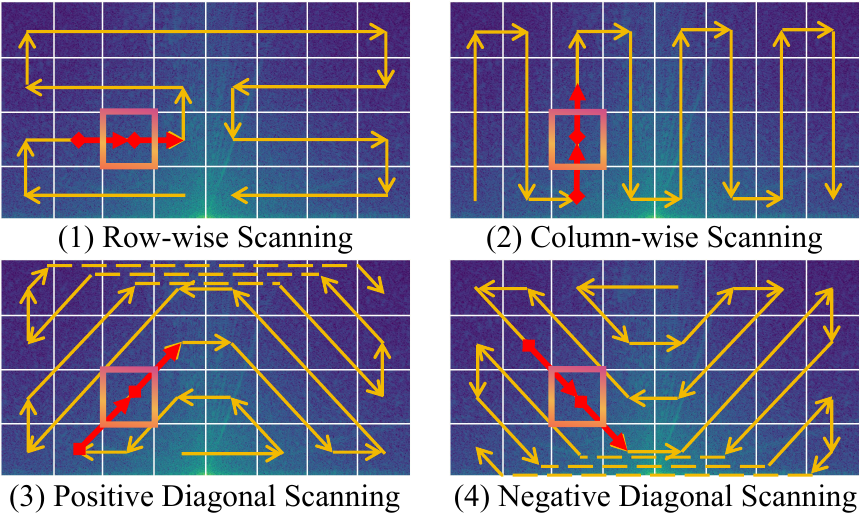}
\vspace{-5pt}
\caption{\textbf{Illustration of our developed Omnidirectional Spectral Scanning (OS-Scan) mechanism,} which incorporates four scanning methods working in Fourier domain: row-wise, column-wise, positive diagonal, and negative diagonal scanning. The red arrows represent the associations between the element highlighted in the box and other elements in the amplitude spectrum.}
\label{fig:osscan}
\vspace{-13pt}
\end{figure}

\subsection{Omnidirectional Spectral State Space Model}
We introduce State Space Model to the exposure correction framework and adapt it to the frequency domain by designing our Omnidirectional Spectral State Space Model (OS-SSM), as illustrated in Figure~\ref{fig:structure}(b).

Given an input feature $\mathbf{X}_\text{in} \in \mathbb{R}^{H\times W\times C}$, we first transform it into $\mathbf{X} \in \mathbb{R}^{H\times W\times D}$ using a linear layer and a LayerNorm~\cite{lei2016layer} operator, then apply a 2D FFT $\mathcal{F}$ to $\mathbf{X}$.
Considering the central symmetry of Fourier spectrum, we retain only half of the spectrum with $\mathcal{A}_h(\mathbf{X}), \mathcal{P}_h(\mathbf{X}) \in \mathbb{R}^{H \times (\frac{W}{2}+1) \times D}$. 
To effectively capture the long-range dependencies in $\mathcal{A}_h(\mathbf{X})$ and $\mathcal{P}_h(\mathbf{X})$, we employ Mamba on amplitude and phase in parallel and propose a novel omnidirectional spectral scanning mechanism (OS-Scan). Each of these Amplitude and Phase Mambas adopt an architecture similar to Vision State-Space Module (VSSM)~\cite{guo2024mambair}, which consists of the sequential operations: \texttt{DWConv} $\rightarrow$ \texttt{Silu} $\rightarrow$ \texttt{OS-Scan} $\rightarrow$ \texttt{S6} $\rightarrow$ \texttt{OS-Merge} $\rightarrow$ \texttt{LayerNorm}.
Different from previous Mamba variants which only use row- and column-wise scanning~\cite{huang2024localmamba, liu2024vmamba, guo2024mambair, shi2024vmambair, hu2024zigma}, 
OS-Scan enhances spectral interactions using positive and negative diagonal scanning with continuous Zig-Zag trajectories~\cite{wallace1992jpeg, hu2024zigma, li2024fouriermamba} to capture more diverse dependencies in $\mathcal{A}_h(\mathbf{X})$ and $\mathcal{P}_h(\mathbf{X})$, as illustrated in Figure~\ref{fig:osscan}. Since the frequencies increase uniformly from the center to both sides, exhibiting a symmetric characteristic, all scanning trajectories are designed to be symmetrical to better leverage this symmetry of the spectrum. 

The scanned sequences are processed by S6~\cite{gu2023mamba} and then reconstructed into 2D feature maps (OS-Merge). 
Following the designs in~\cite{li2024fouriermamba, huang2022fecnet}, we transform the outputs of Amplitude and Phase Mamba back into the spatial domain using the 2D inverse FFT (IFFT) $\mathcal{F}^{-1}$. These transformed features are then modulated by multiplying with SiLU-activated~\cite{elfwing2018sigmoid} input features before applying the LayerNorm and linear layer. 
Overall, the main process of omnidirectional spectral state space model can be expressed as:
\begin{equation}
\begin{aligned}
&\tilde{\mathcal{A}_h}(\mathbf{X})=\text{Amplitude Mamba}(\mathcal{A}_h(\mathbf{X})),\\
&\tilde{\mathcal{P}_h}(\mathbf{X})=\text{Phase Mamba}(\mathcal{P}_h(\mathbf{X})),\\
&\mathbf{X}_\text{modulated} = \mathcal{F}^{-1}(\tilde{\mathcal{A}_h}(\mathbf{X}), \tilde{\mathcal{P}_h}(\mathbf{X})) \odot \text{SiLU}(\mathbf{X}),\\
&\mathbf{X}_\text{out} = \text{Linear}(\text{LayerNorm}(\mathbf{X}_\text{modulated})),
\end{aligned}
\end{equation}
where $\odot$ denotes the element-wise multiplication.

\subsection{Dual-Domain Prior Extractor}
While OS-SSM can correct illumination and recover structures from exposure-error images, image details in severely degraded regions are still too difficult to reconstruct as their high-frequency information is largely lost due to extreme exposure conditions. Such regions require external guidance from well-exposed images to restore the missing details. To achieve this, we use a Dual-Domain Prior Extractor (DDPE) to extract a compact prior $\mathbf{Z} \in \mathbb{R}^{1 \times 1 \times M}$ from $\mathbf{I}_\text{gt}$ and $\mathbf{I}_\text{error}$, injecting it into each OS-SSM to provide correct information for restoring the lost details in these regions.

Specifically, as shown in Figure~\ref{fig:structure}(c), the extraction of DDPE comprises two steps.
The first step is Frequency Feature Extraction, which employs two parallel branches of five residual blocks with $3 \times 3$ convolutions~\cite{he2016deep} to extract features from the amplitude and phase of the processed inputs, which are obtained by downsampling the concated $\mathbf{I}_\text{error}$ and $\mathbf{I}_\text{gt}$ via PixelUnshuffle. The second step is Spatial Feature Refinement, where the amplitude and phase features, transformed back into the spatial domain, are refined using three convolutions, a Global Average Pooling (GAP) operation, and two ReLU-activated~\cite{nair2010rectified} linear layers to produce a compact prior. 
This DDPE design aggregates information from both spatial and frequency domains, enabling better extraction of a useful prior representation compared to previous single-domain designs~\cite{xia2023diffir, wu2023latent, he2023reti}.

After the prior extraction, each OS-SSM uses a linear layer to produce two representations $\mathbf{Z}_1$ and $\mathbf{Z}_2$ from $\mathbf{Z}$, and apply an affine transformation to $\mathbf{X}_\text{out}$ to obtain the prior-guided refined feature $\mathbf{X}_\text{guided}$, which can be formulated as:
\begin{equation}
\begin{aligned}
&\mathbf{Z} = \text{DDPE}(\mathbf{I}_\text{error},\mathbf{I}_\text{gt}),\\
&\mathbf{Z}_1, \mathbf{Z}_2 = \text{Linear}(\mathbf{Z}),\\
&\mathbf{X}_\text{guided} = \mathbf{X}_\text{out} \odot \mathbf{Z}_1 + \mathbf{Z}_2.
\end{aligned}
\end{equation}
It is worth noting that the extraction of $\mathbf{Z}$ is performed only once, while both the spatial sizes of $\mathbf{Z}_1$ and $\mathbf{Z}_2$ are expanded to match that of $\mathbf{X}_\text{out}$ before the pointwise affine transformation denoted by $\odot$ and $+$.

To integrate DDPE into the UNet, we employ an L1 loss $\mathcal{L}_\text{S1}=\lVert \mathbf{I}_\text{gt} - \mathbf{I}_\text{corrected} \rVert_1$ to jointly train them from scratch during the first pretraining stage. This optimizes DDPE to extract effective priors, guiding the UNet to produce corrected images that contain more realistic details than those corrected by a UNet alone, which lacks the capability to restore such details under challenging exposure conditions.

\begin{algorithm}[!t]
\caption{Training Procedure of OSMamba}
\label{alg:osmamba}
\textbf{Input:} Exposure-error images $\mathbf{I}_\text{error}$, ground truth $\mathbf{I}_\text{gt}$. \\
\textbf{Output:} Trained OSMamba. \\
\textbf{Stage 1: Pretraining}
\begin{algorithmic}[1]
\FOR{each training iteration}
\STATE $\mathbf{Z}=\text{DDPE}(\mathbf{I}_\text{error}, \mathbf{I}_\text{gt})$
\STATE $\mathbf{I}_\text{corrected} = \text{UNet}(\mathbf{I}_\text{error}, \mathbf{Z})$
\STATE $\mathcal{L}_\text{S1} = \lVert \mathbf{I}_\text{gt} - \mathbf{I}_\text{corrected} \rVert_1$
\STATE Update the weights of UNet and $\text{DDPE}$.
\ENDFOR
\end{algorithmic}
\textbf{Stage 2: Finetuning}
\begin{algorithmic}[1]
\STATE Freeze the weights of $\text{DDPE}$ as the teacher model.
\STATE Initialize UNet using pretrained weights from stage 1.
\FOR{each training iteration}
\STATE $\mathbf{Z} = \text{DDPE}(\mathbf{I}_\text{error}, \mathbf{I}_\text{gt})$
\STATE $\mathbf{D} = \text{DDPE*}(\mathbf{I}_\text{error})$
\STATE $\hat{\mathbf{Z}}_T \sim \mathcal{N}(\sqrt{\bar{\alpha}_T} \mathbf{Z}, (1-\bar{\alpha}_T)\mathbf{I})$
\FOR{$t=T$ to $1$}
\STATE $\hat{\mathbf{Z}}_{t-1} = \frac{1}{\sqrt{\alpha_t}}\left( \hat{\mathbf{Z}}_t - \frac{1-\alpha_t}{\sqrt{1-\bar{\alpha}_t}}\boldsymbol{\epsilon}_\mathbf{\Theta}(\hat{\mathbf{Z}}_t, t, \mathbf{D}) \right)$
\ENDFOR
\STATE $\mathbf{I}_\text{corrected} = \text{UNet}(\mathbf{I}_\text{error}, \hat{\mathbf{Z}}_0)$
\STATE $\mathcal{L}_\text{S2} = \lVert \mathbf{I}_\text{gt} - \mathbf{I}_\text{corrected} \rVert_1 + \lVert \mathbf{Z} - \hat{\mathbf{Z}}_0 \rVert_1$ 
\STATE Update the weights of UNet, $\boldsymbol{\epsilon}_{\mathbf{\Theta}}$ and $\text{DDPE*}$.
\ENDFOR
\end{algorithmic}
\end{algorithm}

\subsection{Dual-Domain Prior Generator}
Although DDPE can be jointly trained with the UNet to extract priors from $\mathbf{I}_\text{error}$ and $\mathbf{I}_\text{gt}$, it cannot be directly used during deployment since $\mathbf{I}_\text{gt}$ is unavailable in inference. Therefore, a GT-free component is needed that can produce a prior as effective as that from DDPE, but using only $\mathbf{I}_\text{error}$.

To this end, we develop a second fine-tuning stage for OSMamba, as shown in Figure~\ref{fig:structure}(c). Inspired by the great success of knowledge distillation~\cite{xia2022knowledge, gou2021knowledge} and diffusion prior-based methods~\cite{xia2023diffir, wu2023latent, he2023reti}, we freeze the $\text{DDPE}$ trained in stage 1 as a teacher and introduce a Dual-Domain Prior Generator (DDPG) as the student, transferring the knowledge of $\text{DDPE}$ to DDPG. Specifically, we add noise to compact prior $\mathbf{Z}$ extracted by $\text{DDPE}$ to obtain $\hat{\mathbf{Z}}_T$, and design the DDPG as a conditional diffusion model~\cite{saharia2022image, zhang2023adding} that includes a condition extractor $\text{DDPE*}$ and a Denoising Network $\boldsymbol{\epsilon}_{\mathbf{\Theta}}$. The $\text{DDPE*}$ has the same structure as $\text{DDPE}$ except for the input convolution, producing a condition $\mathbf{D} \in \mathbb{R}^{1\times 1\times M}$ from only $\mathbf{I}_\text{error}$. The Denoising Network $\boldsymbol{\epsilon}_{\mathbf{\Theta}}$ consists of two ReLU-activated linear layers. It learns to progressively denoise from a starting point $\hat{\mathbf{Z}}_T$ and condition $\mathbf{D}$ over $T$ steps to generate a diffusion prior $\hat{\mathbf{Z}}_0\in\mathbb{R}^{1\times 1\times M}$ under the noising schedule $\{\alpha_t\}_{t=1}^{T}$ of DDPM~\cite{ho2020denoising, nichol2021improved}. We initialize UNet with the pretrained UNet weights from the first stage and fine-tune it along with the $\text{DDPE*}$ and $\boldsymbol{\epsilon}_{\mathbf{\Theta}}$ using loss $\mathcal{L}_\text{S2}=\lVert \mathbf{I}_\text{gt} - \mathbf{I}_\text{corrected} \rVert_1 + \lVert \mathbf{Z} - \hat{\mathbf{Z}}_0 \rVert_1$, where the second term distills the knowledge of $\text{DDPE}$ into DDPG. The processes of pretraining and finetuning for OSMamba are detailed in Algorithm~\ref{alg:osmamba}.

During inference, we sample a random Gaussian noise $\hat{\mathbf{Z}}_T\sim \mathcal{N}(\mathbf{0},\mathbf{I})$ as the starting point without using $\mathbf{Z}$ to generate the prior $\hat{\mathbf{Z}}_0$, thus eliminating the dependency on $\mathbf{I}_\text{gt}$.

\begin{table*}[!t]
\centering
\resizebox{1\textwidth}{!}{
\setlength{\tabcolsep}{1mm}
\begin{tabular}{l|l|cc|cc|cc|cc|cc|cc|cc}
\shline
              &         & \multicolumn{6}{c|}{MSEC}                                                           & \multicolumn{6}{c|}{SICE}                                                           & \multicolumn{2}{c}{LCDP}    \\ \cline{3-16}
Method        & Source  & \multicolumn{2}{c|}{Under} & \multicolumn{2}{c|}{Over} & \multicolumn{2}{c|}{Average} & \multicolumn{2}{c|}{Under} & \multicolumn{2}{c|}{Over} & \multicolumn{2}{c|}{Average} & \multicolumn{2}{c}{Average} \\ \cline{3-16}
              &         & \multicolumn{1}{c}{PSNR} & \multicolumn{1}{c|}{SSIM}        & \multicolumn{1}{c}{PSNR} & \multicolumn{1}{c|}{SSIM}        & \multicolumn{1}{c}{PSNR} & \multicolumn{1}{c|}{SSIM}         & \multicolumn{1}{c}{PSNR} & \multicolumn{1}{c|}{SSIM}        & \multicolumn{1}{c}{PSNR} & \multicolumn{1}{c|}{SSIM}        & \multicolumn{1}{c}{PSNR} & \multicolumn{1}{c|}{SSIM}         & \multicolumn{1}{c}{PSNR} & \multicolumn{1}{c}{SSIM}         \\ \hline \hline
RetinexNet~\cite{wei2018deep}    & BMVC 2018  & 12.13       & 0.6209      & 10.47      & 0.5953      & 11.14        & 0.6048       & 12.94       & 0.5171      & 12.87      & 0.5252      & 12.90        & 0.5212       & 19.25        & 0.7041       \\
SID~\cite{chen2018learning}      & CVPR 2018 & 19.37 & 0.8103 & 18.83 & 0.8055 & 19.04 & 0.8074 & 19.51 & 0.6635 & 16.79 & 0.6444 & 18.15 & 0.6540 & 21.89 & 0.8082 \\
DRBN~\cite{yang2020fidelity}      & CVPR 2020 & 19.74 & 0.8290 & 19.37 & 0.8321 & 19.52 & 0.8309 & 17.96 & 0.6767 & 17.33 & 0.6828 & 17.65 & 0.6798 & 15.47 & 0.6979 \\
Zero-DCE~\cite{guo2020zero}      & CVPR 2020 & 14.55 & 0.5887 & 10.40 & 0.5142 & 12.06 & 0.5441 & 16.92 & 0.6330 & 7.11 & 0.4292 & 12.02 & 0.5311 & 12.59  & 0.6530 \\
RUAS~\cite{liu2021retinex}          & CVPR 2021  & 13.43       & 0.6807      & 6.39       & 0.4655      & 9.20         & 0.5515       & 16.63       & 0.5589      & 4.54       & 0.3196      & 10.59        & 0.4393       & 13.76        & 0.6060       \\
SCI~\cite{ma2022toward}           & CVPR 2022  & 9.97        & 0.6681      & 5.83      & 0.5190      & 7.49         & 0.5786       & 17.86       & 0.6401      & 4.45       & 0.3629      & 12.49        & 0.5051       & 15.96        & 0.6646       \\
MSEC~\cite{afifi2021msec}          & CVPR 2021  & 20.52       & 0.8129      & 19.79      & 0.8156      & 20.08        & 0.8145       & 19.62       & 0.6512      & 17.59      & 0.6560      & 18.58        & 0.6536       & 20.38        & 0.7800       \\
ENC+SID~\cite{huang2022enc}     & CVPR 2022 & 22.59 & 0.8423 & 22.36 & 0.8519 & 22.45 & 0.8481 & 21.30 & 0.6645 & 19.63 & 0.6941 & 20.47 & 0.6793 & 22.66 & 0.8195 \\
ENC+DRBN~\cite{huang2022enc}    & CVPR 2022  & 22.72       & 0.8544      & 22.11      & 0.8521      & 22.35        & 0.8530       & 21.89       & 0.7071      & 19.09      & 0.7229      & 20.49        & 0.7150       & 23.08         & 0.8302       \\
ECLNet~\cite{huang2022eclnet}        & ACM MM 2022 & 22.37       & 0.8566      & 22.70      & 0.8673      & 22.57        & 0.8631       & 22.05       & 0.6893      & 19.25      & 0.6872      & 20.65        & 0.6861       & 22.44        & 0.8061       \\
FECNet~\cite{huang2022fecnet}        & ECCV 2022  & 22.96       & 0.8598      & 23.22      & 0.8748      & 23.12        & 0.8688       & 22.01       & 0.6737      & 19.91      & 0.6961      & 20.96        & 0.6849       & 22.34        &  0.8038       \\
LCDPNet~\cite{wang2022lcdp}       & ECCV 2022  & 22.35       & \color[HTML]{0500FB}\underline{0.8650}      & 22.17      & 0.8476      & 22.30        & 0.8552       & 17.45       & 0.5622      & 17.04      & 0.6463      & 17.25        & 0.6043       & 23.24        & 0.8420       \\
ERL+FECNet~\cite{huang2023erl} & CVPR 2023 & 23.10 & \color[HTML]{009901}\textit{0.8639} & 23.18 & \color[HTML]{009901}\textit{0.8759} & 23.15 & \color[HTML]{009901}\textit{0.8711} & 22.35 & 0.6671 & 20.10 & 0.6891 & 21.22 & 0.6781 & 22.51 & 0.8153 \\
LACT~\cite{baek2023lact} & ICCV 2023 & \color[HTML]{0500FB}\underline{23.49} & 0.8620 & \color[HTML]{0500FB}\underline{23.68} & 0.8720 & \color[HTML]{0500FB}\underline{23.57} & 0.8690 & 22.35 & \color[HTML]{0500FB}\underline{0.7102} & \color[HTML]{009901}\textit{20.54} & 0.7197 & 21.45 & 0.7150 & 23.69 & \color[HTML]{0500FB}\underline{0.8593} \\
MMHT~\cite{li2023fearless} & ACM MM 2023 & 22.97 & 0.8560 & 23.10 & 0.8709 & 23.05 & 0.8650 & \color[HTML]{009901}\textit{22.55} & \color[HTML]{009901}\textit{0.7090} & \color[HTML]{0500FB}\underline{21.06} & \color[HTML]{009901}\textit{0.7237} & \color[HTML]{0500FB}\underline{21.81} & \color[HTML]{0500FB}\underline{0.7164} & \color[HTML]{009901}\textit{23.72} & 0.8551 \\
CoTF~\cite{li2024real} & CVPR 2024 & \color[HTML]{009901}\textit{23.36} & 0.8630 & \color[HTML]{009901}\textit{23.49} & \color[HTML]{0500FB}\underline{0.8793} & \color[HTML]{009901}\textit{23.44} & \color[HTML]{0500FB}\underline{0.8728} & \color[HTML]{0500FB}\underline{22.90} & 0.7029 & 20.13 & \color[HTML]{FE0000}\textbf{0.7274} & \color[HTML]{009901}\textit{21.51} & \color[HTML]{009901}\textit{0.7151} & \color[HTML]{0500FB}\underline{23.89} & \color[HTML]{009901}\textit{0.8581} \\
\rowcolor{verylightgray} \textbf{OSMamba} & \textbf{Our Proposed} & \color[HTML]{FE0000}\textbf{23.82} & \color[HTML]{FE0000}\textbf{0.8682} & \color[HTML]{FE0000}\textbf{23.75} & \color[HTML]{FE0000}\textbf{0.8823} & \color[HTML]{FE0000}\textbf{23.78} & \color[HTML]{FE0000}\textbf{0.8767} & \color[HTML]{FE0000}\textbf{23.63} & \color[HTML]{FE0000}\textbf{0.7163} & \color[HTML]{FE0000}\textbf{22.01} & \color[HTML]{0500FB}\underline{0.7238} & \color[HTML]{FE0000}\textbf{22.82} & \color[HTML]{FE0000}\textbf{0.7201} & \color[HTML]{FE0000}\textbf{24.53} & \color[HTML]{FE0000}\textbf{0.8773} \\
\shline
\end{tabular}}
\caption{\textbf{Quantitative results of different methods on the MSEC, SICE, and LCDP datasets in terms of PSNR (dB) and SSIM.} A higher metric value indicates better performance. The best result is highlighted in {\color[HTML]{FE0000}\textbf{bold red}}, the second-best in {\color[HTML]{0500FB}\underline{underlined blue}}, and the third-best in {\color[HTML]{009901}\textit{italic green}}. LCDP contains only mixed exposure images, so we do not differentiate between over- or under-exposed cases.}
\label{table:baseline}
\vspace{-5pt}
\end{table*}

\section{Experiment}
\subsection{Experimental Settings}
\noindent \textbf{Datasets.} We evaluate our method on three representative datasets: the multi-exposure dataset MSEC~\cite{afifi2021msec}, SICE~\cite{cai2018learning}, and the mixed exposure dataset LCDP~\cite{wang2022lcdp}. MSEC contains images with two under-exposure and three over-exposure levels for each scene. We use 17,675 images for training and 5,905 for testing. SICE includes both under- and over-exposure levels, with the train-test split following~\cite{huang2022enc}. LCDP contains images with mixed under- and over-exposed regions, consisting of 1,415 image pairs for training and 218 pairs for testing.

\noindent \textbf{Implementation details.} Our method is implemented using the PyTorch framework~\cite{paszke2019pytorch}, with all experiments conducted on NVIDIA RTX 4090 GPUs. The network contains 7.5M parameters. We employ progressive training with image patch sizes [192, 256, 320, 384, 512] and corresponding batch sizes [8, 4, 2, 2, 1]. The learning rate follows a cosine annealing schedule~\cite{loshchilov2016sgdr}, decreasing from $2\times 10^{-4}$ to $1\times 10^{-6}$. We use the Adam optimizer~\cite{diederik2014adam} with parameters $\beta_1 = 0.9$ and $\beta_2 = 0.99$. The training steps for LCDP, MSEC, and SICE in Stage 1 (Pretraining) are 60K, 100K, and 80K, respectively, followed by 20K joint finetuning steps in Stage 2 (Finetuning). The channel numbers $C$ and $M$ of feature and prior are set to 36 and 256, respectively. For the diffusion model in DDPG, we use a linear noise schedule with $\alpha_1=0.9$, $\alpha_T=0.01$, and $T=4$.

\begin{figure*}[!t]
\centering
\includegraphics[width=1\textwidth]{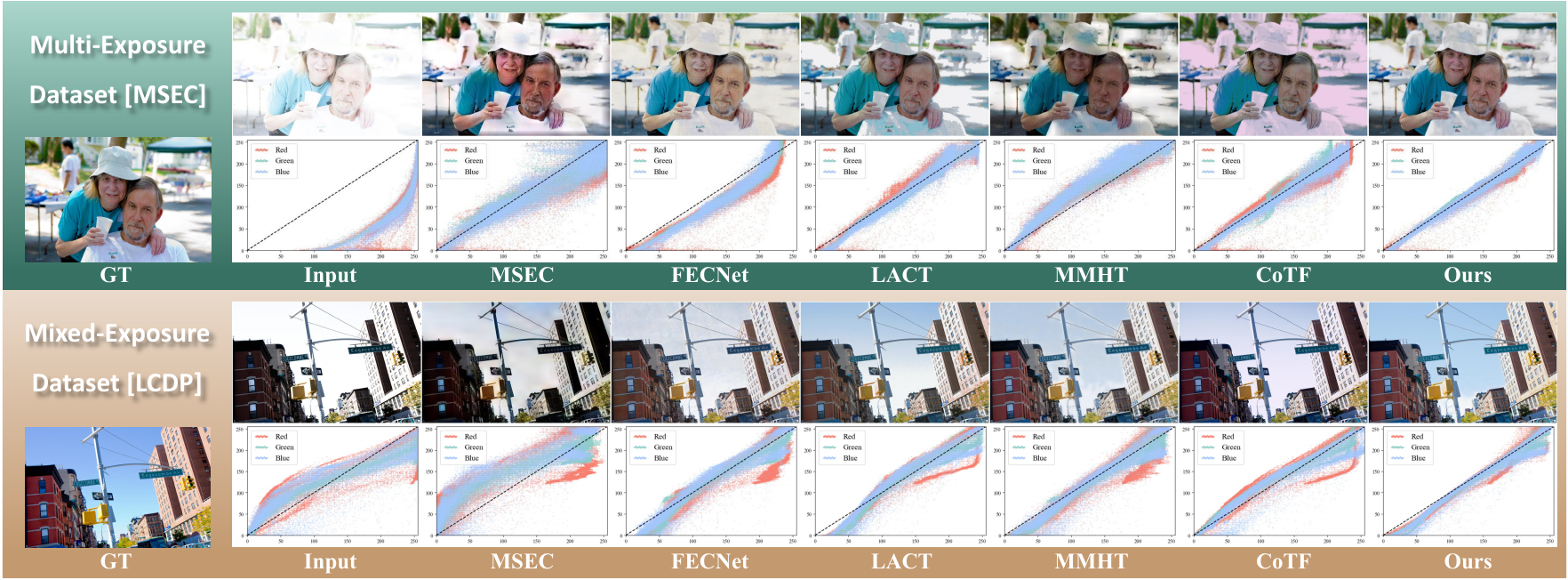}
\caption{\textbf{Visual comparison of our method against previous state-of-the-art approaches.} The first row displays comparisons on the multi-exposure dataset MSEC, while the second row presents comparisons on the mixed-exposure dataset LCDP.}
\label{fig:showimg}
\end{figure*}

\subsection{Comparison with State-of-the-Art Methods}
\noindent \textbf{Quantitative Comparisons.} Table~\ref{table:baseline} presents a quantitative comparison among our proposed method and state-of-the-art approaches. On the multi-exposure dataset MSEC, our method outperforms existing methods in both under- and over-exposure conditions, demonstrating robust correction capabilities across various exposure errors. Similarly, on the SICE dataset, our method shows a significant improvement of 1.01dB in PSNR over the second-best method. For the mixed exposure dataset LCDP, our method exhibits notable advantages, with a 0.64 dB increase in PSNR compared to COTF and a 0.0353 improvement in SSIM over LCDPNet. These results underscore the effectiveness of our method in correcting complex mixed exposure inputs.

\noindent \textbf{Qualitative Comparisons.} Figure~\ref{fig:showimg} illustrates the qualitative comparison between our method and other approaches, accompanied by RGB scatter plots mapping the corrected image to the ground truth below each result. The closer the scatter points are to the diagonal, the better the correction effect. For the MSEC dataset, we showcase a severely overexposed image. The correction results from MSEC, LACT, and MMHT exhibit significant detail distortion in the table and shed on the right, as well as in the hat and clothing. The outputs of FECNet and CoTF suffer from imbalances in the illumination and structure, along with color tone deviations. In contrast, our method not only corrects illumination and recovers structure well but also restores details in severely degraded areas with natural colors. For the LCDP dataset, we present an image with severe mixed over- and under-exposure. The correction results from MSEC, LACT, and MMHT show notable detail distortions in the sky and on the walls of buildings on both sides. FECNet and CoTF fail to accurately recover the original structure and illumination of the sky. Our method effectively corrects the colors in severely degraded areas while minimizing distortions.

\subsection{Ablation Study and Discussion}
This subsection conducts ablation study on LCDP to verify the efficacy of OS-SSM and Dual-Domain Prior Generator, which are our two main methodological contributions.

\begin{figure*}[!t]
\centering
\includegraphics[width=0.98\textwidth]{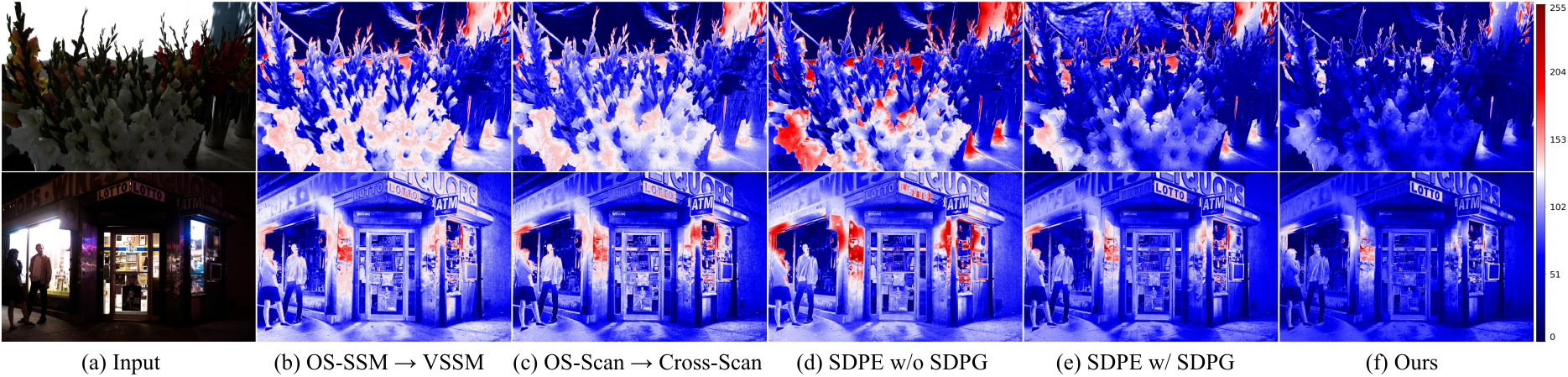} 
\caption{\textbf{Visual comparison of error maps between correction results from different ablation methods and the ground truth.} ``$\rightarrow$'' denotes the replacement of the left module with the right module. From left to right, the five images correspond to setting (a) in Table~\ref{table:block-setting}, setting (e) in Table~\ref{table:scanning-setting}, settings (b) and (c) in Table~\ref{table:prior-setting}, and our proposed method, respectively.}
\label{fig:abexpfig}
\end{figure*}

\begin{table}[!t]
\centering
\resizebox{0.65\columnwidth}{!}{
\setlength{\tabcolsep}{1.5mm}
\begin{tabular}{cccc}
\toprule
Setting & Model Type & PSNR & SSIM \\
\midrule
(a) & VSSM & 24.27 & 0.8748 \\
(b) & Attention & 24.31 & 0.8762 \\
\textbf{OSMamba} & \textbf{OS-SSM} & \textbf{24.53} & \textbf{0.8773} \\
\bottomrule
\end{tabular}}
\caption{\textbf{Ablation study on the model type used in OS-SSB.} We replace the OS-SSM in each OS-SSB with either the Vision State-Space Module (VSSM)~\cite{guo2024mambair} or an Attention module~\cite{zamir2022restormer}.}
\label{table:block-setting}
\vspace{-10pt}
\end{table}

\noindent \textbf{Effectiveness of Omnidirectional Spectral SSM.} In Table~\ref{table:block-setting}, we replace our proposed OS-SSM with two different models, while keeping the FFNs and a comparable number of parameters. Firstly, in setting (a), we replace OS-SSM with the Vision State-Space Module (VSSM)~\cite{guo2024mambair}, which adopts a Cross-Scan mechanism~\cite{liu2024vmamba} in the spatial domain. This substitution results in a 0.26dB decrease in PSNR. The visual comparison between Figure~\ref{fig:abexpfig}(f) and (b) highlights the advantages of our frequency domain-based OS-SSM over the spatial domain-based VSSM. In particular, Figure~\ref{fig:abexpfig}(b) shows significant structure distortions and errors in illumination estimation in blossoms and walls, compared to Figure~\ref{fig:abexpfig}(f). Secondly, in setting (b), we replace OS-SSM with an Attention module~\cite{zamir2022restormer}, which leads to a 0.22dB drop in PSNR. These results confirm the superiority of our proposed OS-SSM.

\begin{table}[!t]
\centering
\resizebox{0.95\columnwidth}{!}{
\setlength{\tabcolsep}{1.5mm}
\begin{tabular}{ccccccc}
\toprule
\multirow{2}{*}{Setting} & \multicolumn{2}{c}{Cross-Scan~\cite{liu2024vmamba}} & \multicolumn{2}{c}{OS-Scan} & \multirow{2}{*}{PSNR} & \multirow{2}{*}{SSIM} \\
\cmidrule(lr){2-3} \cmidrule(lr){4-5}
& No.1-2 & No.3-4 & No.1-2 & No.3-4 & & \\
\midrule
(a) & $\checkmark$ & $\times$ & $\times$ & $\times$ & 24.30 & 0.8757 \\
(b) & $\times$ & $\checkmark$ & $\times$ & $\times$ & 24.32 & 0.8760 \\
(c) & $\times$ & $\times$ & $\checkmark$ & $\times$ & 24.34 & 0.8758 \\
(d) & $\times$ & $\times$ & $\times$ & $\checkmark$ & 24.37 & 0.8769 \\
(e) & $\checkmark$ & $\checkmark$ & $\times$ & $\times$ & 24.35 & 0.8761 \\
\textbf{OSMamba} & $\times$ & $\times$ & $\checkmark$ & $\checkmark$ & \textbf{24.53} & \textbf{0.8773} \\
\bottomrule
\end{tabular}}
\caption{\textbf{Ablation study on the scanning method used in OS-SSM.} Cross-Scan~\cite{liu2024vmamba} No.1-2 represents horizontal and vertical scanning starting from the top-left, while No.3-4 represents the reverse directions. OS-Scan No.1-2 corresponds to row-wise and column-wise scanning, while No.3-4 denote positive and negative diagonal scanning.}
\label{table:scanning-setting}
\vspace{-10pt}
\end{table}

Table~\ref{table:scanning-setting} compares five distinct scanning configurations. Specifically, we use Cross-Scan~\cite{liu2024vmamba} as a comparison to OS-Scan. Comparing settings (a), (b), and (e), Cross-Scan No.1-4 fails to achieve further improvement over No.1-2 and No.3-4, indicating that merely relying on reverse direction scanning cannot establish more effective correlations in the frequency spectrum.
Comparing settings (c), (d), and OSMamba, OS-Scan No.1-4 achieves further improvement over No.1-2 and No.3-4. Comparing setting (e) and OSMamba, OS-Scan No.1-4 gains a 0.18 dB improvement over Cross-Scan No.1-4. This demonstrates that row-wise and column-wise scanning complemented by positive and negative diagonal scanning forms an effective synergy, achieving better results than conventional scanning through richer spectral correlations. Figure~\ref{fig:abexpfig}(f) shows superior structure recovery, particularly in areas such as bouquets and clothing compared to Figure~\ref{fig:abexpfig}(c).

\begin{table}[!t]
\centering
\resizebox{0.95\columnwidth}{!}{
\setlength{\tabcolsep}{1.5mm}
\begin{tabular}{ccccccc}
\toprule
\multirow{2}{*}{Setting} & \multicolumn{2}{c}{Spatial-Domain} & \multicolumn{2}{c}{Dual-Domain} & \multirow{2}{*}{PSNR} & \multirow{2}{*}{SSIM} \\
\cmidrule(lr){2-3} \cmidrule(lr){4-5}
 & SDPE & SDPG & DDPE & DDPG & & \\
\midrule
(a) & $\times$ & $\times$ & $\times$ & $\times$ & 23.72 & 0.8561 \\
(b) & $\checkmark$ & $\times$ & $\times$ & $\times$ & 24.01 & 0.8687 \\
(c) & $\checkmark$ & $\checkmark$ & $\times$ & $\times$ & 24.32 & 0.8771 \\
\textbf{OSMamba} & $\times$ & $\times$ & $\checkmark$ & $\checkmark$ & \textbf{24.53} & \textbf{0.8773} \\
\bottomrule
\end{tabular}}
\caption{\textbf{Ablation study on dual-domain prior generator.} We compare four different settings for producing the prior $\mathbf{Z}$. SDPE and SDPG denote the Spatial-Domain Prior Extractor and Generator, respectively, which eliminate the FFT and IFFT operations used in our DDPE and DDPG, and instead employ ten residual blocks~\cite{he2016deep} to extract deep features solely in the spatial domain.}
\label{table:prior-setting}
\vspace{-5pt}
\end{table}

\noindent \textbf{Effectiveness of Dual-Domain Prior Generator.}  
Table~\ref{table:prior-setting} compares three OSMamba variants with different settings of producing the prior $\mathbf{Z}$. To be concrete, firstly, in setting (c), we replace the frequency feature extraction in $\text{DDPE}$ and $\text{DDPE*}$ with single-domain spatial feature extraction by eliminating the FFT and IFFT operations, instead using ten residual blocks that operate solely in the spatial domain. This brings a 0.21dB PSNR decrease. The visual comparison between Figure~\ref{fig:abexpfig}(e) and (f) highlights the importance of aggregating information from both spatial and frequency domains for effective exposure correction. In Figure~\ref{fig:abexpfig}(e), using only spatial domain processing increases the correction error in challenging regions such as local textures on flower buds and road surfaces, compared to our dual-domain design in Figure~\ref{fig:abexpfig}(f). Secondly, in setting (b), we remove the latent diffusion model and only use an extractor to get the prior $\mathbf{Z}$ from $\mathbf{I}_\text{error}$, resulting in a 0.52dB PSNR drop. Visual comparisons between Figure~\ref{fig:abexpfig}(d) and (e) show that eliminating the generative diffusion model increases distortions in regions with flower petals, walls, and store signs. Finally, setting (a) eliminates the use of prior completely, relying solely on a UNet for exposure correction, which leads to significant decreases of 0.81 dB in PSNR and 0.0212 in SSIM. These results comprehensively demonstrate the effectiveness of prior guidance, generative latent diffusion model, and dual-domain design for DDPE.

\begin{figure}[!t]
\centering
\includegraphics[width=\columnwidth]{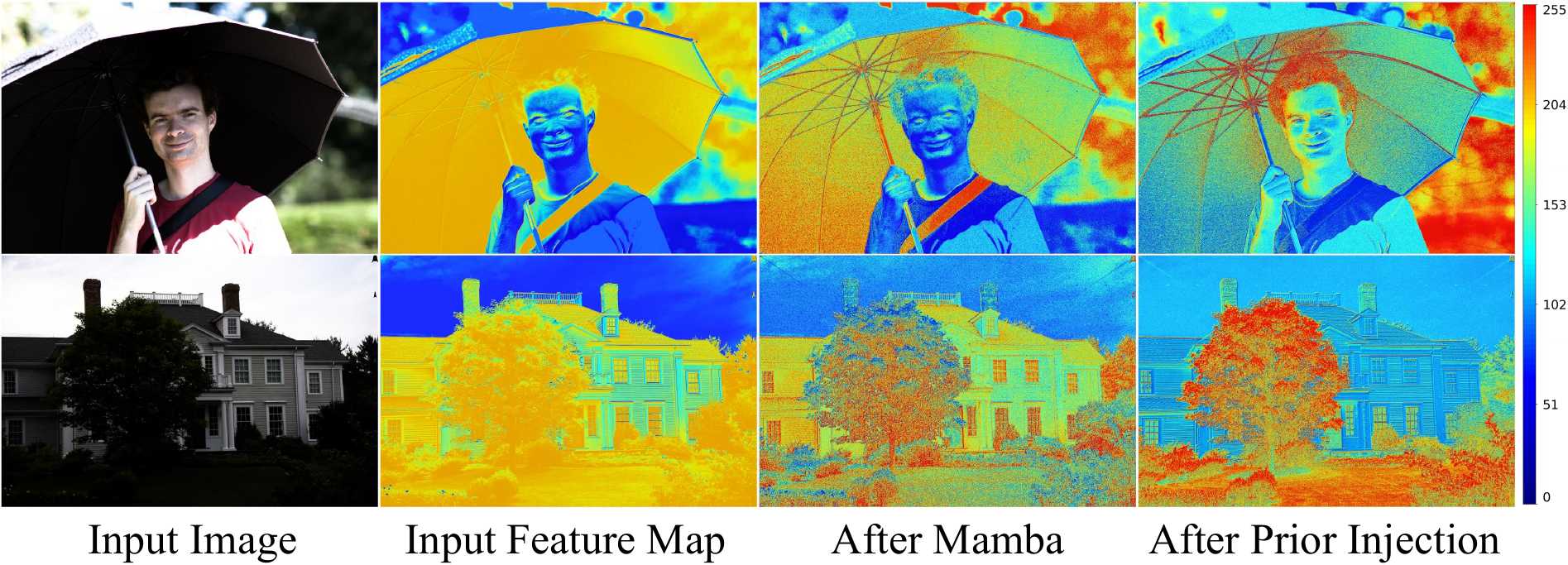} 
\caption{\textbf{Visual comparison of the intermediate feature maps at different positions of the first OS-SSM in our trained OSMamba.} We use heatmaps to visualize the input features (2nd column), the results after processing by Amplitude and Phase Mambas (3rd column), and the results after prior injection (4th column).}
\label{fig:heatmap}
\vspace{-5pt}
\end{figure}

\subsection{Visual Analysis}
Figure~\ref{fig:heatmap} presents the visualization of feature maps at different positions within the first OS-SSM in our trained OSMamba. This visualization provides valuable insights into the functionality of the modules in OS-SSM. Firstly, we observe that the features processed by our Amplitude and Phase Mambas with OS-Scan effectively capture the structures of the umbrella and buildings, which were not apparent in the input features. Secondly, more intricate details in textured areas such as the face, umbrella ribs, sky, grass, and trees are further enhanced by the injection of generative diffusion prior. We conjecture that these visual results can indicate the effectiveness of OS-Scan and prior injection in progressively refining structure and retrieving lost details.

\section{Conclusion}
This paper proposes OSMamba, a novel exposure correction method leveraging our developed OS-Scan mechanism to capture long-range dependencies across multiple directions in both the amplitude and phase spectra, effectively correcting illumination and recovering structures. In addition, a DDPG module is designed to efficiently guide the correction network by aggregating information from spatial and frequency domains to restore lost details. Experimental results demonstrate that OSMamba outperforms previous state-of-the-art methods, validating the effectiveness of our OS-Scan and DDPG design in enhancing exposure correction performance. In the future, we plan to explore the potential of OSMamba in HDR image reconstruction.

\noindent \textbf{Acknowledgments.} This work was supported in part by National Natural Science Foundation of China (No. 62372016), Guangdong Provincial Key Laboratory of Ultra High Definition Immersive Media Technology (No. 2024B1212010006) and Shenzhen General Research Project under Grant JCYJ20220531093215035.

\small
\bibliographystyle{ieeenat_fullname}
\bibliography{ref}

\end{document}